\documentclass[aps,prd,twocolumn]{revtex4-1}
\usepackage{epsfig}

\begin{document}
\author{Zhu-fang Cui$^{1,4}$, Chao Shi$^{2,4}$, Wei-min Sun$^{2,3,4}$, Yong-long Wang$^{1,5,6}$, and Hong-shi Zong$^{2,3,4,}$}\email[]{zonghs@chenwang.nju.edu.cn}
\address{$^{1}$ Key Laboratory of Modern Acoustics, MOE, Institute of Acoustics, and Department of Physics, Nanjing University, Nanjing 210093, China}
\address{$^{2}$ Department of Physics, Nanjing University, Nanjing 210093, China}
\address{$^{3}$ Joint Center for Particle, Nuclear Physics and Cosmology, Nanjing 210093, China}
\address{$^{4}$ State Key Laboratory of Theoretical Physics, Institute of Theoretical Physics, CAS, Beijing, 100190, China}
\address{$^{5}$ Department of Physics, School of Science, Linyi University, Linyi 276005, P. R. China}
\address{$^{6}$ Center for Theoretical Physics, Massachusetts Institute of Technology, Cambridge, MA02139, USA}

\title{The Wigner Solution and QCD Phase Transitions in a Modified PNJL Model}
\begin{abstract}
By employing some modification to the widely used two-flavor Polyakov-loop extended Nambu-Jona-Lasinio (PNJL) model, we discuss the Wigner solution of the quark gap equation at finite temperature and zero quark chemical potential beyond the chiral limit, and then try to explore its influences on the chiral and deconfinement phase transitions of QCD at finite temperature and zero chemical potential. The discovery of the coexistence of the Nambu and the Wigner solutions of the quark gap equation with nonzero current quark mass at zero temperature and zero chemical potential, as well as their evolutions with temperature is very interesting for the studies of the phase transitions of QCD. According to our results, the chiral phase transition might be of first order (while the deconfinement phase transition is still a crossover, as in the normal PNJL model), and the corresponding phase transition temperature is lower than that of the deconfinement phase transition, instead of coinciding with each other, which are not the same as the conclusions obtained from the normal PNJL model. In addition, we also discuss the sensibility of our final results on the choice of model parameters.

\bigskip


\bigskip


\end{abstract}
\maketitle

In the nonperturbative regime of Quantum Chromodynamics (QCD), chiral symmetry breaking and quark color confinement are of great importance and continuous interests for studying the QCD phase diagram. However, their relation is not yet clarified directly from the first principles of QCD. Generally speaking, color confinement indicates chiral symmetry breaking, while the reverse is not necessarily true. How these two phenomena are related to each other and whether (and/or under which conditions) these two transitions coincide when the temperature and/or quark chemical potential grow larger have been speculated and discussed by many people via many a model, for example, see Refs. \cite{K83,B83,W11,M04,S07,C09,A08,FM09,S10,M10,F11,FL11,B11,B12,P12,BF12}. Strictly speaking, chiral and deconfinement phase transitions only occur in opposite sectors in QCD. Chiral symmetry is an exact global symmetry only when the current quark mass $m_{\rm q}$ is zero (the chiral limit). In the low-temperature phase (hadronic phase, often referred to as Nambu-Goldstone phase or Nambu phase), this symmetry is spontaneously broken, and as a consequence there exist $N_f^2 - 1$ pseudoscalar Nambu-Goldstone bosons, meanwhile the QCD vacuum hosts a chiral condensate (two quark condensate) $\langle{\bar q}q \rangle$, which acts as an order parameter for chiral phase transition. However, the $Z(3)$ center symmetry associated with the color confinement is exact only in the limit of pure gauge QCD, which means $m_{\rm q}\rightarrow \infty$, and so of course is too far from our real world. In the high-temperature, deconfinement phase (Wigner phase, where the quark-gluon plasma, or QGP, is expected to appear) of QCD, this symmetry is spontaneously broken, where the Polyakov loop \cite{P78}, which is related to the heavy quark free energy, can serve as an order parameter for deconfinement phase transition. For the case of finite physical quark mass, neither the quark condensate nor the Polyakov loop is a good order parameter.

It is generally believed that with increasing temperature or baryon number density, strongly interacting matterwill undergo a phase transition from the hadronic matter to the QGP which is expected to appear in the ultrarelativistic heavy ion collisions or the inner core of compact stars. As for the nature of the phase transitions, a standard scenario may favor a crossover at small chemical potential, both for the chiral phase transition and the deconfinement phase transition, and then turning into a first order chiral transition for larger chemical potential at a critical end point (CEP). This picture is consistent with Lattice QCD simulations and various QCD-inspired models. The search for the CEP is also one of the main motivations in the experiments. However, on the theoretical side there is still an ambiguity, not only for the location of CEP, but also for whether this standard scenario is correct. For example, in Ref. \cite{B13} the authors argue that there is no CEP, since the transition is a crossover in the whole phase diagram; in Ref. \cite{CN10} the authors also think that there is no CEP, but a Lifshitz point instead; the authors of Refs. \cite{M02,B03,H06} consider there may be two CEPs; while the authors of Refs. \cite{B05,A10} find that the CEP and triple point are possible to coincide with each other, due to existence of another phases (namely, color superconducting or quarkyonic matter) at low temperature and high density. Even in the case of one CEP, there are still uncertainties on the position of the CEP (see, for example, Ref. \cite{S06}). Unfortunately, lattice Monte Carlo simulations cannot be used to resolve this issue due to the notorious sign problem, so the calculations based on effective theories of QCD are also irreplaceable nowadays. The purpose of this work is to introduce some modification to the widely used two-flavor Polyakov-loop extended
Nambu-Jona-Lasinio (PNJL) model and try to explore its influences on the chiral and deconfinement phase transitions of QCD at finite temperature and zero chemical potential.

Usually, Nambu and Wigner phases are described respectively by two different solutions of the quark gap equation. Although the existence of those two solutions is generally accepted in the chiral limit, it is generally believed that the quark gap equation only has the Nambu-Goldstone solution beyond the chiral limit, whereas the Wigner solution disappears. This is in fact not compatible with the current studies of QCD phase transitions.  The authors of Ref. \cite{Z05} doubted this issue firstly, and discussed whether the quark gap equation has a Wigner solution in the case of nonzero current quark mass, and hereafter, the authors of Refs. \cite{C07,W07,FN09,W12} investigated this further. However, this problem has not been solved satisfactorily until now. In this work, by employing some modification to the widely used PNJL model \cite{F04,R06}, and based on the study in Refs. \cite{J12,CEPJC13}, we discuss the Wigner solution at finite temperature and zero quark chemical potential when the current quark mass $m_{\rm q}$ is nonzero. As will be discussed later, the discovery of the coexistence of the Nambu and the Wigner solutions of the quark gap equation beyond the chiral limit when temperature and chemical potential are both zero, along with their evolutions with temperature is very interesting for the studies of the phase transitions of QCD. Moreover, we display the calculated result of the chiral and deconfinement phase transitions in the case of zero chemical potential and finite temperature, and furthermore, we also make some discussions on the effects of varying the weight factor of the influence of the quark propagator to the gluon propagator.

In the normal PNJL model \cite{F04,R06}, the following generalized Lagrangian density is introduced, with quarks coupled to a (spatially constant) temporal background gauge field representing Polyakov loop dynamics (here we take the number of flavors $N_{\rm f} = 2$ and the number of colors $N_{\rm c}=3$):
\begin{eqnarray}
\mathcal{L}_{\rm PNJL}&=~&\mathcal{L}_0+G \mathcal{L}_{\rm I}+\mathcal{U}\nonumber\\
&=~&\bar{\psi}\left(i\gamma_{\mu}D^{\mu}-\hat{m}_{\rm q}\right)\psi+G\left[\left(\bar{\psi}\psi\right)^2+\left(\bar{\psi}i\gamma_5 \mathbf{\tau}\psi \right)^2\right]\nonumber\\
 &&-\mathcal{U}\left(\Phi[A],\bar{\Phi}[A],T\right), \label{eq1}
\end{eqnarray}
where $\psi=\left(\psi_{\rm u},\psi_{\rm d}\right)^{\rm T}$ is the quark field and
\begin{equation}
D^{\mu}=\partial^\mu-i A^\mu~~~~~~\rm{and}~~~~~~A^\mu=\delta_{\mu0}A^0~~. \label{eq2}
\end{equation}
The gauge coupling constant $g$ is conveniently absorbed into the definition of $A^\mu(x) = g {\cal A}^\mu_a(x){\lambda_a/ 2}$, with ${\cal A}^\mu_a$ being the SU(3) gauge field and $\lambda_a$ being the Gell-Mann matrices. The mass matrix is $\hat{m}_{\rm q} ={\rm diag}(m_{\rm u}, m_{\rm d})$. When working in the limit of exact isospin symmetry, people often take $m_{\rm u} = m_{\rm d} \equiv m_{\rm q}$. A local, chirally symmetric scalar-pseudoscalar four-point interaction of the quark fields is introduced with an effective coupling strength $G$. It should be noted that, since $G$ is taken to be a constant in the normal (P)NJL model, it is the same in different phases (even in the chiral limit, where in principle there should be no DCSB in the Wigner phase), and does not change when the temperature and/or quark chemical potential vary. However, as we will explain later, the coupling strength should not only differ for different phases (especially, it cannot cause DCSB for the Wigner phase in the chiral limit), but also has temperature and chemical potential dependence.

The Polyakov loop $L$ is an $SU(N_{\rm c})$ matrix in color space,
\begin{equation}
L\left(\bf{x}\right)=\mathcal{P}\exp\left[i\int_{0}^{\beta}d\tau\,A_4\left(\bf{x},\tau\right)\right], \label{eq3}
\end{equation}
with $\cal P$ denoting the path ordering operation, $\beta = 1/T$ is the inverse temperature and $A_4 = i A^0$. $\mathcal{U}(\Phi,\bar{\Phi},T)$ is the effective potential expressed in terms of the traced Polyakov loop and its (charge) conjugate (throughout our calculation both will be treated as classical field variables),
\begin{equation}
\Phi=({\rm Tr_c}\,L)/N_{\rm c},~~~~\bar{\Phi}=({\rm Tr_c}\,L^{\dagger})/N_{\rm c}. \label{eq4}
\end{equation}
In the pure gauge sector, one would have $\Phi=0$ below a critical temperature $T_0$, and $\Phi\rightarrow1$ in the limit $T\rightarrow\infty$.
The form proposed in Ref. \cite{R06} by comparison with Lattice QCD will be adopted throughout our calculation (all the parameters we used in this work are summarized in Table 1 and Table 2):
\begin{equation}
{\mathcal{U}\left(\Phi,\bar{\Phi},T\right)\over T^4} =-{b_2\left(T\right)\over 2}\bar{\Phi} \Phi- {b_3\over 6}\left(\Phi^3+ {\bar{\Phi}}^3\right)+ {b_4\over 4}\left(\bar{\Phi} \Phi\right)^2, \label{eq5}
\end{equation}
with
\begin{equation}
b_2\left(T\right)=a_0+a_1\left(\frac{T_0}{T}\right)+a_2\left(\frac{T_0}{T}\right)^2+a_3\left(\frac{T_0}{T}\right)^3~. \label{eq6}
\end{equation}
$T_0$ is the critical value for deconfinement appearing in the pure gauge sector. A typical value 270 MeV is employed, as used in Refs. \cite{F04,R06}. In the limiting case
mainly discussed in this paper, where quark chemical potential $\mu =0$, one would find that $\Phi=\bar{\Phi}$ and then Eq. (\ref{eq5}) can be simplified.

As usual, the effective quark mass $M$ can be determined via the self-consistent gap equation:
\begin{equation}
M=m_{\rm q}-2G\langle\bar{\psi}\psi\rangle. \label{eq7}
\end{equation}
The quark condensate is defined as
\begin{eqnarray}
\langle\bar\psi\psi\rangle&=&-\int \frac{\mathrm{d}^4p}{(2\pi)^4}\mbox{Tr}[G(p)],\label{eq8}
\end{eqnarray}
where $G(p)$ is the dressed quark propagator, and the trace is to be taken in color, flavor, and Dirac space. Strictly speaking, this quantity is ultraviolet divergent and such divergence cannot be eliminated by the usual renormalization procedure \cite{Z03}. However, in (P)NJL model one can impose the cutoff $\Lambda$ to regularize the integral, and then this problem is avoided. In the normal (P)NJL model, one would obtain $M=325$ MeV from Eq. (\ref{eq7}) as the Nambu solution, and no Wigner solution exists beyond the chiral limit. After some algebra, the thermodynamic potential per unit volume in the mean field approximation can be obtained \cite{R06}:
\begin{eqnarray}
\Omega&=~&{\cal U}\left(\Phi,\bar{\Phi},T\right)+G\langle\bar{\psi}\psi\rangle^2- 6N_{\rm f}\int^\Lambda\frac{\mathrm{d}^3p}{\left(2\pi\right)^3}{E_p}\nonumber\\
&-&2N_f\,T\int\frac{\mathrm{d}^3p}{\left(2\pi\right)^3} \left[\ln f1+\ln f2\right], \label{eq9}
\end{eqnarray}
where
\begin{eqnarray*}
f1=&&~1+\mathrm{e}^{-\frac{3\left(E_p-\mu\right)}{T}}+3\left(\Phi+\bar{\Phi}\mathrm{e}^{-\frac{\left(E_p-\mu\right)}{T}}\right)\mathrm{e}^{-\frac{\left(E_p-\mu\right)}{T}},\\
f2=&&~1+\mathrm{e}^{-\frac{3\left(E_p+\mu\right)}{T}}+3\left(\bar{\Phi}+\Phi\mathrm{e}^{-\frac{\left(E_p+\mu\right)}{T}}\right)\mathrm{e}^{-\frac{\left(E_p+\mu\right)}{T}},
\end{eqnarray*}
$E_p=\sqrt{{\bf p}^2+M^2}$ is the quark quasi-particle energy and $\Lambda$ is the three-momentum cutoff from the normal (P)NJL model. The cut-off $\Lambda$ is only imposed on the first integration (zero-point energy). The second integration, which represents the finite temperature contribution, has a natural cut-off in itself just specified by the temperature. It is explicitly shown that when $T=0$ the Polyakov loop and the quark sector decouple.

\begin{figure}
\includegraphics[width=8cm]{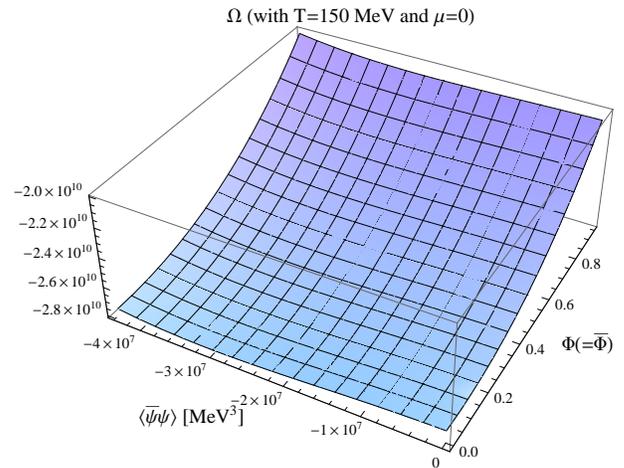}
\caption{The thermodynamic potential density $\Omega$ as a function of the chiral condensate $\langle\bar{\psi}\psi\rangle$ and the Polyakov loop $\Phi(=\bar{\Phi})$, where the temperature $T$ is fixed to be 150 MeV with zero chemical potential $\mu$.}\label{fig1}
\end{figure}

\begin{figure}
\includegraphics[width=8cm]{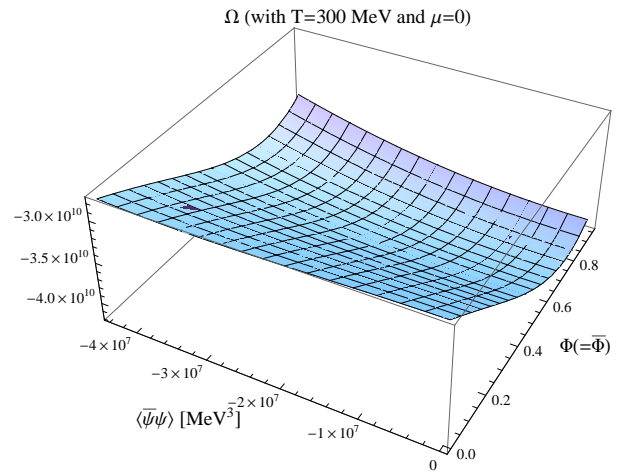}
\caption{The thermodynamic potential density $\Omega$ as a function of the chiral condensate $\langle\bar{\psi}\psi\rangle$ and the Polyakov loop $\Phi(=\bar{\Phi})$, where the temperature $T$ is fixed to be 300 MeV with zero chemical potential $\mu$.}\label{fig2}
\end{figure}

Now we would like to display the thermodynamic potential density $\Omega$ as a function of the chiral condensate and the Polyakov loop in Fig. \ref{fig1} and Fig. \ref{fig2} for two different temperatures respectively: $T=150$ MeV (below $T_0$) and $T=300$ MeV (above $T_0$), while the chemical potential is fixed to be zero (then $\Phi=\bar{\Phi}$, which is easy to understand just from the expression of $\Omega$, Eq. (\ref{eq9}), and is shown clearly in the actual calculations).

From the comparison between Fig. \ref{fig1} and Fig. \ref{fig2} one can notice that, the dependence of $\Omega$ on $\Phi(=\bar{\Phi})$ is much larger than that on $\langle\bar{\psi}\psi\rangle$. Our calculation shows that this phenomenon holds at least for the whole temperature range discussed in this work. When the temperature is lower than some critical value, as illuminated in Fig. \ref{fig1}, the minimal value of the thermodynamic potential density $\Omega$ is located in the region where the absolute value of chiral condensate is large and the value of Polyakov loop is very small, which just corresponds to the Nambu solution of the gap equation, Eq. (\ref{eq7}). As the temperature increases, the location of the minimum for $\Omega$ will continuously goes to the region with smaller chiral condensate and larger Polyakov loop. For the temperatures above some critical value, as revealed in Fig. \ref{fig2}, the corresponding effective quark mass $M$ will be much smaller than the normal Nambu one, and is close to the current quark mass $m_{\rm q}$, which is just the Wigner solution of gap equation. These confirm that the Lagrangian (\ref{eq1}) can satisfactorily describe chiral symmetry restoration and quark deconfinement simultaneously to a certain degree, even beyond the chiral limit. Moreover, we want to stress that, as illustrated by the plots, for a given temperature the thermodynamic potential has only one minimum, which means that the gap equation only has one solution: either the Nambu one or the Wigner one, they do not coexist.

The exact values of $\langle\bar{\psi}\psi\rangle$, $\Phi$, and $\bar{\Phi}$ for a given $(T,~\mu)$ can be obtained by solving the following equations in a self-consistent way:
\begin{equation}
{\partial\Omega\over\partial\langle\bar{\psi}\psi\rangle} = {\partial\Omega\over\partial\Phi} = {\partial\Omega\over\partial \bar{\Phi}} = 0 ~. \label{eq11}
\end{equation}
One would find that when one imposes the condition $\mu=0$, $\Phi$ and $\bar{\Phi}$ will equal to each other for all values of $T$.

It is well known that the quark propagator and the gluon propagator satisfy their respective Dyson-Schwinger Equations (DSEs), and they are coupled with each other \cite{R94,R00}. As a result of the quark propagators in Nambu and Wigner phase being so different, the corresponding gluon propagators in these two phases should be different, too. Just as pointed out and discussed in Refs. \cite{Z05,J12,CEPJC13}, the differences between the vacuum of Nambu phase and  Wigner phase can be characterized by the quark condensate (which is associated with the spontaneous breaking of chiral symmetry). Therefore, the gluon propagator would be different due to different quark condensate in these two phases (obviously, in the normal (P)NJL model this has never been considered). Similar discussions have already been performed and verified in quantum electrodynamics for 2+1 dimensions (QED$_3$, which has many features similar to QCD, such as spontaneous chiral symmetry breaking in the massless fermion limit and confinement, and thus can serve as a toy model of QCD), for the fermion and the photon propagators \cite{F06}.

At present it is impossible to calculate the influence of the quark propagator to the gluon propagator from the first principle of QCD. So one has to resort to various nonperturbative QCD models to express them phenomenologically. Over the past few years, considerable progress has been made in the framework of the QCD sum rule \cite{Shifman}, which provides a successful description of various nonperturbative aspects of strong interaction physics at both zero and finite temperature. We naturally expect that it might provide some useful clue to the study of the nonperturbative contribution of the quark propagator to the model gluon propagator.

From the plane wave method of QCD sum rule \cite{SumR1}, the non-perturbative part of a Green function is defined as the difference between the full Green function (which is unknown) and the perturbative part. The condensates are then identified with the various moments of the nonperturbative Green function. So the most general form of the ``nonperturbative'' gluon propagator should be
\begin{eqnarray}
D_{\mu\nu}^{npert}\equiv D_{\mu\nu}^{full}-D_{\mu\nu}^{pert}\equiv c_1\langle\bar{\psi}\psi\rangle+c_2\langle G^{\mu\nu}G_{\mu\nu}\rangle+\cdot\cdot\cdot, \nonumber
\end{eqnarray}
where $\langle\bar{\psi}\psi\rangle$ and $\langle G^{\mu\nu}G_{\mu\nu}\rangle$ are the two-quark condensate and gluon condensate, respectively, the coefficients $c_1$ and $c_2$ can be calculated using the QCD sum rule approach~\cite{S89,SumR2}, and the ellipsis represents the contribution from other condensates, e.g., the mixed quark-gluon condensate. Among all the condensates, the two-quark condensate (a nonvanishing value of which will signal the DCSB in the chiral limit) has the lowest dimension, and is generally believed to be the most important condensate in the QCD sum rule approach. Hence, in this work we will pick out the contribution of the two-quark condensate separately, and the contribution from other condensates will be added into the perturbative gluon propagator. In the normal (P)NJL model, this is equivalent to modifying the coupling constant $G$ in the following way:
\begin{equation}
G\rightarrow G1+G2 \langle\bar{\psi}\psi\rangle.\label{eq10}
\end{equation}
Physically, it is well known that QCD has a non-trivial vacuum structure. One way to characterize this structure is by means of various vacuum condensates. These condensates are also essential for describing the strong interaction physics using the QCD sum rule method. So, when gluons propagate in the nonperturbative vacuum of QCD, they will certainly be affected by these condensates~\cite{S89,SumR2}. Just as discussed above, among all the condensates, the two-quark condensate is generally believed to be the most important one in describing the nonperturbative vacuum of QCD. Hence in this work we pick it out (and the effects of all the other condensates are simplified into the first term $G1$ of Eq. (\ref{eq10})) to study its qualitative influences on the gluon propagator, and then on the chiral and deconfinement phase transitions of QCD. Therefore, $G1$ is an effective coupling strength that reflects all the other contributions besides the part proportional to the two-quark condensate to the gluon propagator, and is considered to be the same in both Nambu and Wigner phases; while $G2 \langle\bar{\psi}\psi\rangle$ is different in Nambu and Wigner phases. And briefly speaking, once all the parameters are chosen, we can regard $G2$ as an effective coupling strength that reflects the weight factor of the influence of the quark propagator to the gluon propagator. We hope that by such a simple model one can capture the essential physics of QCD phase transitions.

\begin{figure}
\includegraphics[width=8cm]{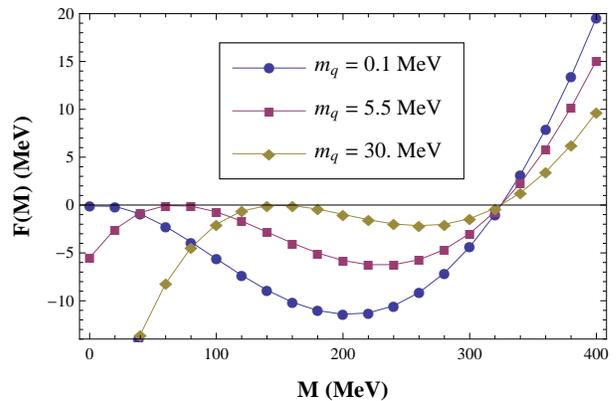}
\caption{Solutions of the gap equation when the chiral condensate is separated to characterize the differences between the vacuum of Nambu phase and that of Wigner phase.}\label{fig3}
\end{figure}

Now let us turn to the determination of the model parameters in this work. The way to fix the values of the new parameters $G1$ and $G2$ in this work is illustrated in Fig. \ref{fig3}, where
\begin{equation}
F(M)=M-m_{\rm q}-2(G1+G2\langle\bar{\psi}\psi\rangle)\langle\bar{\psi}\psi\rangle.\label{eqfm}
\end{equation}
is the gap equation.

The value of $G1+G2 \langle\bar{\psi}\psi\rangle$ for $M=325$ MeV with temperature and chemical potential being both zero is fixed to be $5.04\times10^{-6}$ MeV$^{-2}$, which equals the value of $G_{\rm NJL}$ in the normal (P)NJL model. We find that only when $G1=3.16\times 10^{-6}$ MeV$^{-2}$ and $G2=-5.91\times 10^{-14}$ MeV$^{-5}$, can one get the result for $m_{\rm q}=5.5$ MeV shown in Fig. \ref{fig3}, where the equation $F(M)=0$ has and only has two solutions. One solution, $M=325$ MeV, is the ordinary Nambu solution; the other one, which is much smaller that it (approximately 68 MeV, about half of the mass of a pion, and will continuously tends to zero when $m_{\rm q}$ approaches zero, as can be seen clearly from Fig. \ref{fig3}), could be identified as the Wigner solution that describes the perturbative dressing effect in the case $m_{\rm q}\neq 0$.

\begin{figure}
\includegraphics[width=8cm]{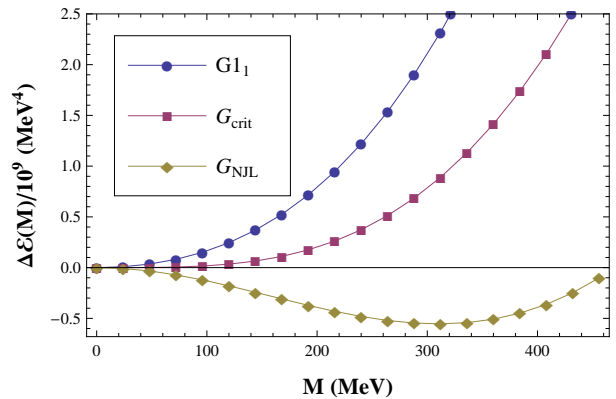}
\caption{The vacuum energy density difference between Nambu phase and Wigner phase (in the chiral limit) for different coupling constants. For details, please see the text.}\label{GDCSB}
\end{figure}

In the normal (P)NJL model, it is well-known that only when the coupling constant $G$ is above a critical value will chiral symmetry breaking happen. Specifically, in the case of chiral limit, there should be no DCSB in the Wigner phase, which means that the coupling constant in the Wigner phase
should be smaller than this critical coupling constant. In our model, in the case of chiral limit and in Wigner phase, $G2 \langle\bar{\psi}\psi\rangle\equiv0$ and the effective coupling constant $G\equiv G_1$. Just as shown above, in principle, $G_1$ should be smaller than the critical coupling constant.
Then, it is natural to ask whether this requirement is fulfilled in our model. In order to confirm this, here it is necessary and interesting to plot the curve of the vacuum energy density difference between Nambu phase and Wigner phase in the chiral limit, namely,
\begin{eqnarray}
\Delta\mathcal{E}~&=~&\mathcal{E}(M)_{\rm Nambu}-\mathcal{E}(M=0)_{\rm Wigner}\nonumber\\
&=&\frac{M^2}{4G}-\frac{3}{4\pi^2}[\Lambda\sqrt{M^2+\Lambda^2}(M^2+\Lambda^2)\nonumber\\
&&-M^4{\rm sinh}^{-1}(\frac{\Lambda}{M})-2\Lambda^4], \label{deltaE}
\end{eqnarray}
as shown in Fig. \ref{GDCSB}. For comparison, we show three cases of the coupling constant, where $G1_1$ is just $3.16\times 10^{-6}$ MeV$^{-2}$, as chosen above, and there is indeed no DCSB; $G_{\rm crit}=3.88\times 10^{-6}$ MeV$^{-2}$ is a critical value, at which DCSB begins to appear; and $G_{\rm NJL}$ is the coupling constant in the normal (P)NJL model, where clear DCSB is illustrated. Moreover, it should be noted that not only the $G1$ above but also the two cases of $G1$ as will be discussed in Table 3 are smaller than $G_{\rm crit}$, so that this can also be regarded as a self-consistency check of our model.

Strictly speaking, since the coupling contains information from gluons, in principle $G1$ and $G2$ should depend on the Polyakov loop $L$, too. Nevertheless, observing that $L$ is essentially the temporal component of gluons, $G1$ and $G2$ would not be affected by $L$ only, therefore the qualitative results would not change. So, we will simply neglect any possible $L$ dependence, and assume that $G1$ and $G2$ include all the information from gluons, as the way people treat $G$ in the normal PNJL model \cite{F04,R06}. Then, all the parameters used in this work are listed in Table 1 (for the Polyakov potential part) and Table 2 (for the NJL model part).

\begin{table}
\begin{center}
Table 1: Parameter set used in our work for the Polyakov loop potential (\ref{eq5}) and (\ref{eq6}). All parameters are taken from Ref. \cite{R06}.\\
\begin{tabular}{p{1.2cm} p{1.2cm} p{1.2cm} p{1.2cm} p{1.2cm}p{1.2cm}}
\hline
$a_0$&$a_1$&$a_2$&$a_3$&$b_3$&$b_4$\\
\hline
6.75&-1.95&2.625&-7.44&0.75&7.5\\
\hline \label{tb1}
\end{tabular}
\end{center}
\end{table}

\begin{table}
\begin{center}
Table 2. Parameter set used in our work for the NJL model part of the effective Lagrangian (\ref{eq1})\\
\begin{tabular}{p{1.6cm} p{1.6cm} p{2.1cm} p{2.1cm}}\hline
$m_{\rm q}$ & $\Lambda$ & $G1$ & $G2$ \\

[MeV] & [MeV] & [MeV$^{-2}$] & [MeV$^{-5}$]\\
\hline
5.5&651&$3.16\times 10^{-6}$&$-5.91\times 10^{-14}$\\
\hline
\end{tabular}
\label{tb2}
\end{center}
\end{table}

\begin{figure}
\includegraphics[width=8cm]{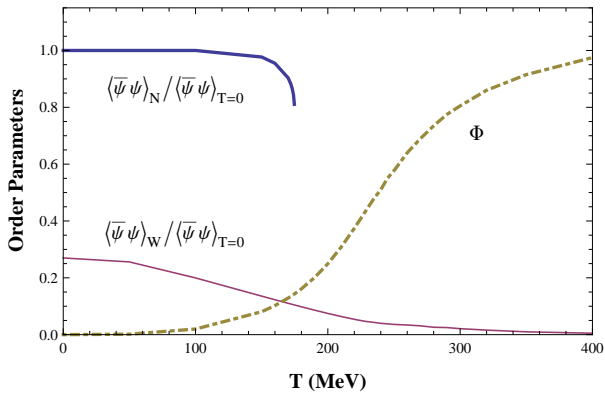}
\caption{Scaled chiral condensate of Nambu phase and Wigner phase together with the Polyakov loop $\Phi(=\bar{\Phi})$, as functions of temperature at zero chemical potential.}\label{fig4}
\end{figure}

After making the replacement of Eq. (\ref{eq10}), we then calculate the temperature dependence of the chiral condensate and the Polyakov loop, with the chemical potential fixed to be zero, too. The results are plotted in Fig. \ref{fig4}.

Generally speaking, since the relation between the NJL model and QCD remains somewhat obscure, accordingly the qualitative results are often more valuable than the quantitative ones. Here it is interesting to compare our results with those of the normal PNJL model, such as Fig. 4 of Ref. \cite{R06}. We find that thanks to the introduction of quarks coupled to both $\langle\bar{\psi}\psi\rangle$ and $\Phi$ fields, the first order deconfinement phase transition seen in pure-gauge Lattice QCD is now a continuous crossover, as found in Refs. \cite{F04,R06}. However, our results show that the chiral phase transition which is indicated by the chiral condensate is obviously of first order, and it does not coincide with the deconfinement phase transition, which is indicated by the Polyakov loop. This is somewhat different from the normal PNJL model. The value of the critical temperature $T_c$ for the first order chiral phase transition (i.e., the temperature at which the Nambu solution disappears) here is about 175 MeV, which is in accordance with the data of two-flavor Lattice QCD, $T_c=173\pm 8$ MeV \cite{K02}, and even the same as the result of Ref. \cite{G11}, meanwhile is much smaller than the temperature of deconfinement. These results show that the chiral phase transition might happen earlier than the deconfinement phase transition, which is qualitatively the same as the result found in the Lattice QCD studies of the Wuppertal-Budapest collaboration, where the critical temperature for the chiral restoration is about 25 MeV lower than the deconfinement one \cite{A06,A09,B10}. And more interestingly, once the difference of the gluon propagators (or in other words, the difference of vacua) between Nambu phase and Wigner phase is taken into consideration via the chiral condensate, we can see that the Nambu solution and Wigner solution would coexist below $T_c$, which is very interesting in the studies of QCD phase transitions, and has never been found in the normal (P)NJL model. The plot shows that the effective masses of both Nambu phase and Wigner phase will decrease as $T$ increases, which means that the dressing effect of quarks becomes weaker and weaker.

For the chiral phase transition with two-flavor quarks, the results above are qualitatively different from previous results in the original PNJL model. When we do not take into account the feedback of the quark condensate to the gluon propagator, our model will reduce to the original PNJL model. Here, a natural question arises: is our treatment reasonable?  This question can be answered from three aspects. Firstly, just as discussed above, using different gluon propagators in different phases is a requirement of QCD, and the treatment in this work can also ensure that there would not be DSCB for the Wigner phase in the chiral limit; Secondly, in the DSEs approach, the bag constant is identified with the pressure difference between Nambu phase and Wigner phase \cite{R94,R00}. However, according to the the usual point of view in the literature, only in the case of chiral limit does the quark gap equation has both the Nambu and the Wigner solutions simultaneously. In other words, in the usual (P)NJL model, only in the chiral limit can one define the bag constant. Nevertheless, in the real world, the current quark mass is nonzero, and the bag models (such as the famous MIT bag model) are constructed for this case, where the bag constant plays an important role. In our work, since the coexistence of the Nambu and the Wigner solutions are founded beyond the chiral limit at zero temperature and zero chemical potential, one can then define the bag constant in this case; Last but not least, in principle, the coupling strength should not only be distinct in different phases, but also vary when temperature and/or chemical potential change. Nevertheless, this is still an open problem, especially in the nonperturbative regime of QCD. Yet in our model setup the coupling strength would change naturally, since the chiral condensate is temperature and chemical potential dependent. Moreover, it is known that in Lattice QCD calculations people employs the functional integral method and determines the phase transition temperature by studying the quark-number susceptibility or the scalar susceptibility, etc. (see, for example, Ref. \cite{AE06}). Consequently, one cannot find the metastable state, or more specifically, the Wigner solution at low temperature. In this case, in Lattice QCD calculations people cannot adopt the pressure difference of these two phases to study the partial restoration of chiral symmetry at finite temperature, whereas in our model study both the true stable and the metastable states can be found simultaneously. In this sense, the model studies of continuum field theory may also provide some hints to the Lattice QCD calculations.

\begin{figure}
\includegraphics[width=8cm]{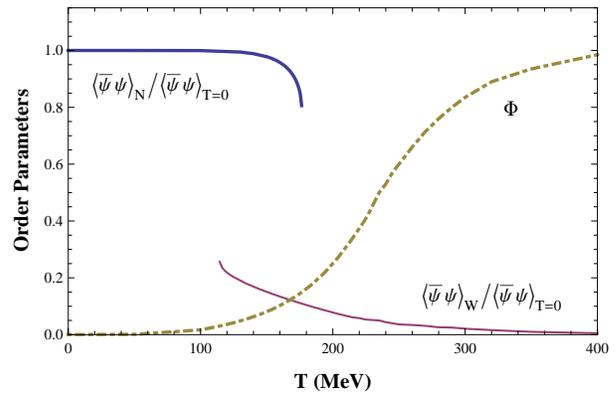}
\caption{Scaled chiral condensate of Nambu phase and Wigner phase together with the Polyakov loop $\Phi(=\bar{\Phi})$, as functions of temperature at zero chemical potential for the case (I) of Table 3.}\label{fig5}
\end{figure}

\begin{figure}
\includegraphics[width=8cm]{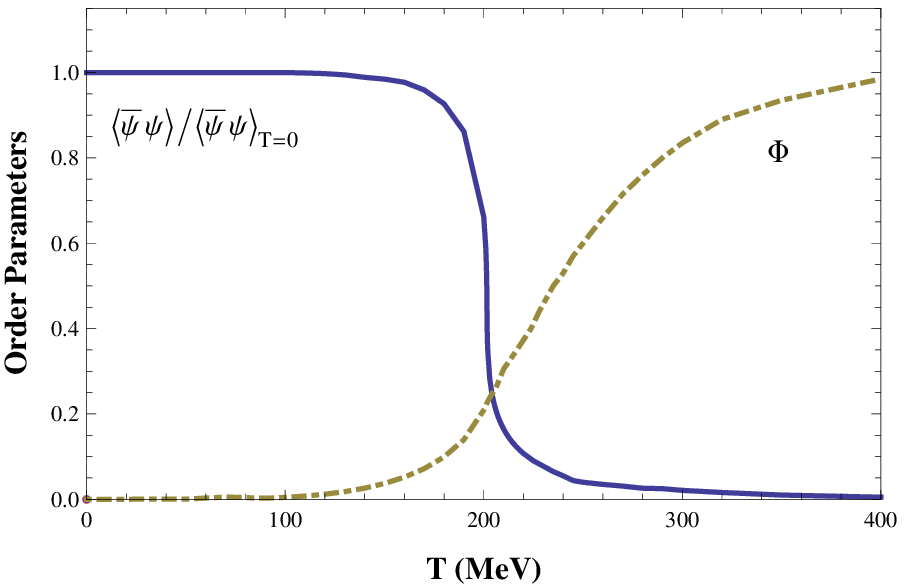}
\caption{Scaled chiral condensate of Nambu phase and Wigner phase together with the Polyakov loop $\Phi(=\bar{\Phi})$, as functions of temperature at zero chemical potential for the case (II) of Table 3.}\label{fig6}
\end{figure}

Just as explained above, the value of $G1+G2\langle \bar{\psi}\psi\rangle$ in our model is fixed to be the coupling constant in the normal (P)NJL model. Now, it is interesting to change the relative weight of $G1$ and $G2\langle \bar{\psi}\psi\rangle$ (while all the other parameters are fixed as before) to see their influence on the results. For instance, in Fig. \ref{fig5} and Fig. \ref{fig6}, we show respectively the two cases of different parameter choices in Table 3.

\begin{table}
\begin{center}
Table 3: Different parameter choices of $G1$ and $G2$.\\
\begin{tabular}{p{2.4cm} p{2.4cm} p{2.4cm}}
\hline
&$G1$&$G2$\\
&[MeV$^{-2}$] & [MeV$^{-5}$]\\
\hline
Case (I)&$3.21\times 10^{-6}$&$-5.77\times 10^{-14}$\\
\hline
Case (II)&$3.61\times 10^{-6}$&$-4.49\times 10^{-14}$\\
\hline \label{tb3}
\end{tabular}
\end{center}
\end{table}

As far as we know, as was mentioned above, due to the difficulty of determining how the gluon propagator is affected by the quark condensate from the first principles of QCD, there is hardly any discussion on this issue, especially in the nonperturbative region. So, here we would like to continue our discussions with larger $G1$ and correspondingly smaller $G2$ (this is understood in the sense of the absolute value of the parameters, and similarly hereinafter), which means that the influence of the quark propagator to the gluon propagator is evaluated to be weaker. We are not interested in the cases where $G1$ is small while the corresponding $G2$ is larger, because in that cases there is no qualitative change in both the Nambu solution and the Wigner solution, nevertheless there would appear another metastable solution, which in our opinion is nonphysical, just as in some cases studied in Ref. \cite{W12}.

As a result of the increase of $G1$ and the corresponding decrease of $G2$, the Wigner solution of the gap equation may not appear at lower temperatures, but begins to coexist with the Nambu solution at some critical temperature, just as illustrated obviously in Fig. \ref{fig5}, case (I) of Table 3. Qualitatively, this is very similar to the discoveries of another recent work of our group, Ref. \cite{J13}, which used quite distinct theoretical tools. Although one is for the case of zero chemical potential and finite temperature, whereas the other is for the case of zero temperature and nonzero chemical potential, we think that the coexistence of the Nambu solution and the Wigner solution of the quark gap equation might be an interesting physical phenomenon which has never been found before, rather than a coincidence or just a mathematical result of the calculations. At the same time, the curve for the crossover of the deconfinement phase transition moves slightly towards the direction of lower temperature, while its shape is basically unchanged. This shows once again that the changes of the chiral properties of the system might not have any obvious qualitative influence on its deconfinement nature. If we continue to take larger $G1$ and correspondingly smaller $G2$, the critical temperatures at which the Nambu solution disappears and the Wigner solution begin to appear would both increase, but the range of region that they coexist would decrease more rapidly. Then, at some critical value of $G1$ (and correspondingly for $G2$), which we take as case (II) of Table 3, the Nambu solution and the Wigner solution would converge with each other, and the first order chiral phase transition discovered above is now a crossover with a very sharp slope, as shown in Fig. \ref{fig6}. Then for even larger $G1$ and correspondingly even smaller $G2$, which means the influence of the quark condensate to the gluon propagator becomes even weaker, the two curves will continue to approach each other, meanwhile the crossover of the chiral phase transition becomes smoother and smoother. Then at the very last, the results of the normal PNJL model and the widely accepted Lattice QCD calculations will be perfectly repeated, as expected. At last, we want to point out that our quantitative results are sensitive to the parameters adopted (such as the critical temperature $T_0$), which is also discussed in Ref. \cite{R06,C10}. However, the qualitative conclusions drawn from our results would not change.

In summary, we have given the model setup of a widely used two-flavor chiral effective model with Polyakov loop dynamics (PNJL model) and discussed the pattern of the solutions of the quark gap equation beyond the chiral limit. Then, we studied the Wigner solution of the quark gap equation with nonzero current quark mass in the case of finite temperature and zero chemical potential by introducing some modification to the normal PNJL model. Usually, people think that the quark gap equation does not have the Wigner solution beyond the chiral limit. However, when we pick out the two-quark condensate effect and investigate its influence on the gluon propagator, the outcome shows that the Wigner solution may coexist with the Nambu solution at nonzero current quark mass. This discovery is very interesting in the studies of both the chiral and  deconfinement phase transitions of QCD. Based on this, we further discuss the chiral and deconfinement phase transitions of QCD at finite temperature and zero chemical potential using the modified two-flavor PNJL model. Our results show that the influence of the Polyakov loop on the thermodynamical potential is much larger than that from the quark condensate, and the Nambu solution would disappear at sufficiently high temperature, rather than a crossover as many people found in the normal PNJL model and the Lattice QCD results. The critical temperature we obtain is about 175 MeV, which is in accordance with the data of two-flavor Lattice QCD. Moreover, according to our results, the chiral phase transition might happen earlier than the deconfinement phase transition, instead of coinciding with each other, which is not the same as the result of the normal PNJL model, but qualitatively the same as the result found in the Lattice QCD studies of the Wuppertal-Budapest collaboration. However, further discussions show that the weight factor of the influence of the quark propagator on the gluon propagator may be crucial for one to draw some reliable conclusions, since it is very difficult to clarify this from the first principles of QCD. For smaller weight of the influence of the quark condensate to the gluon propagator, the coexistence region of the Wigner solution with the Nambu solution may decrease and even disappear, and the first order chiral phase transition found above may degenerate to a widely accepted crossover. However, all the modifications do not show obvious impact on the deconfinement phase transition. These qualitative conclusions obtained in our work do not change with the different choices of the parameters.

\acknowledgments
This work is supported in part by the National Natural Science Foundation of China (under Grant 11275097, 10935001, 11274166 and 11075075), the National Basic Research Program of China (under Grant 2012CB921504) and the Research Fund for the Doctoral Program of Higher Education (under Grant No 2012009111002).

\end{document}